\documentstyle[12pt]{article}

\begin{document}

\author{S.A.Pol'shin\thanks{
E-mail: itl593@online.kharkov.ua} \\
{\normalsize {\it Department of Physics, Kharkov State University, }}\\
{\normalsize {\it Svobody Sq., 4, 310077, Kharkov, Ukraine}}}
\title{Massless fields over $R^1 \times H^3$ space-time and coherent states
for the Lorentz group}
\date{}
\maketitle

\begin{abstract}
The solutions of the arbitrary-spin massless wave equations over ${\bf R}^1
\times H^3$ space are obtained using the generalized coherent states for the
Lorentz group. The use of these solutions for the construction of
invariant propagators of quantized massless fields with an arbitrary spin
over the ${\bf R}^1 \times H^3$ space is considered. The expression for the
scalar propagator is obtained in the explicit form.
\end{abstract}

The method of generalized coherent states (CS) successfully applied for
various physical problems~\cite{coher1} is a  natural way of
describing  the quantized fields over symmetric spaces too.
In the previous paper\cite{dS-PLB} we applied this method
  for the description of quantum fields in the Sitter space.
Here we wish to apply this method to the massless fields over the  ${\bf R}
^{1}\times H^{3}$ space.

To obtain the wave equations over this space in a convenient form,
at first we  write down the Bargmann-Wigner equations for an arbitrary-spin
particle in the Minkowski space in the form of four equations for the
$2s+1$-component spinor (or, equivalently, for some polynomial). We show
that only two of these equations are mutually independent and their
covariant generalization yields the consistent wave equations over the ${\bf
R} ^{1}\times H^{3}$ space. To obtain the solutions of these wave equations,
we introduce two CS systems for the Lorentz group. The first (scalar)
system is determined by a spin zero infinite-dimensional representations of
the Lorentz group and corresponds to the coset space $H^{3}$. The second
(spinor)  system corresponds to  finite-dimensional representations of
the Lorentz group and to the coset space $SO(3,1)/\{$the little Lorentz
group of a light-like vector\}. The scalar CS system multiplied by
$e^{i\omega t}$ gives  the solution of the Klein-Gordon equation for the
conformally-coupled massless scalar field over the ${\bf R}^{1}\times H^{3}$
space. In turn, the solutions of arbitrary-spin wave equations  are presented
by the product of the scalar CS with the imaginary shift of the mass and
spinor CS multiplied by $e^{i\omega t}$. This allows us to construct the
quantized fields over the ${\bf R}^{1}\times H^{3}$ space with the
propagators being invariant under the ${\bf R}^{1}\otimes SO(3,1)$ group.
Here we consider the spin zero case only and show that the corresponding
propagator has a standard form containing the $\delta $-function.

The three-dimensional hyperbolic space $H^{3}$ is the hyperboloid of a
radius $R$ in a four-dimensional Lorentzian space with the fictitious fourth
coordinate equal to $R\kappa _{{\bf x}}$, where $\kappa _{{\bf x}}=(1+{\bf x}
^{2}/R^{2})^{1/2}.$ The symmetry group of the $H^{3}$ space is $SO(3,1)$ and
its irreducible representations ${(j_{+},j_{-})}$ are well-known~\cite{49}.
They are determined by  two complex numbers $j_{+}$ and $j_{-}$ which
correspond to the  eigenvalues of Casimir operators of  two $so(3)$
subalgebras of the $so(3,1)$ algebra. For the case when $j_{+}$ and $j_{-}$
are integer or half-integer the corresponding representation is
finite-dimensional and non-unitary. In the opposite case it is
infinite-dimensional and unitary and may be characterized by the spin
$s=0,1/2,\ldots $ and the frequency $\omega \in {\bf R}$; then the weights are
equal to
\begin{equation}
j_{+}=-\frac{i\omega R+s+1}{2}\quad ,\quad j_{-}=\frac{-i\omega R+s-1}{2}.
\label{weights}
\end{equation}
Let us take $x^{i}$ as a set of coordinates over the $H^{3}$ space; then
the metric over ${\bf R}^{1}\times H^{3}$ is
\[
g^{ij}=-\delta ^{ij}-\frac{x^{i}x^{j}}{R^{2}}\quad ,\quad g^{00}=1\quad
,\quad g^{0i}=0.
\]
The Klein-Gordon equation for the conformally coupled massless field with
spin zero over the ${\bf R}^{1}\times H^{3}$ space has the form
\begin{equation}
(\Box +R^{-2})\psi =0\quad ,\quad \Box \equiv (-\det g_{\mu \nu
})^{-1/2}\partial _{\mu }((-\det g_{\mu \nu })^{1/2}g^{\mu \nu }\partial
_{\nu }).  \label{4. 11a}
\end{equation}
It is shown in~\cite{III} that at $\psi \sim e^{i\omega t}$ the above equation
corresponds to the irreducible representation of the Lorentz group with the
weights~(\ref{weights}) for $s=0$. The solutions of Eq.~(\ref{4. 11a}) may be
presented in the form of "plane waves" analogous to that in the de Sitter
space considered in~\cite{dS-PLB}:
\begin{equation}
\psi ({\bf x},t)=e^{i\omega t}\phi _{{\bf q}}({\bf x};i\omega R-1)\quad
,\quad \phi _{{\bf q}}({\bf x};\sigma )=\left( \kappa _{{\bf x}}-\frac{{\bf
qx}}{R}\right) ^{\sigma },  \label{spin0}
\end{equation}
where ${\bf q}^{2}=1$.  According to~\cite{coher1}  the functions $\phi
_{{\bf q}}({\bf x} ;\sigma )$ are the CS for some representation of
the Lorentz group.  For the case $\sigma =i\omega R-1$ it is irreducible
representation for  spin zero particles.

The wave equations for massless particles with an
arbitrary spin over the Minkowski space have the form~\cite{40}
\begin{equation}
(s\eta ^{\mu \nu }+S^{\mu \nu })\partial _{\nu }\psi =0,  \label{bargwig5}
\end{equation}
where $S_{\mu \nu }$ are the generators of the corresponding
finite-dimensional representation of the Lorentz group. According to
\cite{40}, we consider the realization of the generators $S_{\mu \nu }$ as
the differential operators acting on the variables $z,\bar{z}$.
Then the wave function $\psi $ should be a polynomial in $z$ and
$\bar{z}$ of the power equal or less than $2j_{+}$ and $2j_{-}$, respectively.
We restrict ourselves by the case of particles with the positive helicity;
then the generators $S_{\mu \nu }$ compose the representation ${(s,0)}$ and
have the form
\begin{eqnarray}
S_{ik} &=&-i\varepsilon _{ikl}J_{l}\quad ,\quad S^{0i}=J_{i},  \nonumber \\
J_{+} &=&-\frac{\partial }{\partial z}\quad ,\quad J_{-}=z^{2}\frac{\partial
}{\partial z}-2sz\quad ,\quad J_{3}=s-z\frac{\partial }{\partial z}.
\nonumber
\end{eqnarray}
Here we introduce the indices $+,-$ as $a_{\pm }=a^{1}\pm ia^{2}$ and $J_{i}
$ are the generators of the spin $s$ representation of the $SO(3)$ group.
Putting $\mu =i,0$ in Eqs.~(\ref{bargwig5}) we obtain that they are
equivalent to
\begin{eqnarray}
(s\partial _{0}+J_{i}\partial _{i})\psi  &=&0,  \label{bargwig1a} \\
V_{ik}\partial _{k}\psi  &\equiv &(s\delta
_{ik}-s^{-1}J_{i}J_{k}+i\varepsilon _{ikl}J_{l})\partial _{k}\psi =0.
\label{bargwig2a}
\end{eqnarray}
It is easy to show that
\begin{eqnarray}
V_{-i} &=&-z^{2}V_{+i}\quad ,\quad V_{3i}=zV_{+i}\quad ,\quad V_{++}=-s^{-1}
\frac{d^{2}}{dz^{2}},  \label{bargwig5a} \\
V_{+-} &=&s^{-1}z^{2}\frac{d^{2}}{dz^{2}}+\frac{4s-2}{s}J_{3}\quad ,\quad
V_{+3}=\frac{2s-1}{s}\frac{d}{dz}-s^{-1}z\frac{d^{2}}{dz^{2}}.  \nonumber
\end{eqnarray}
As a result, only one equation of Eqs.~(\ref{bargwig2a}) is independent;
let us choose this equation as $V_{+i}\partial _{i}\psi =0$. Computing the
Weyl tensor of the ${\bf R}^{1}\times H^{3}$ space we can show that it
vanishes and this space is conformally-flat. Then, by the virtue of the
conformal invariance, the consistent ${\bf R}^{1}\times H^{3}$ wave
equations for the massless particle with an arbitrary spin may be obtained
from Eqs.~(\ref{bargwig5}) by replacing $\partial _{\mu }$ by the covariant
derivative
\begin{equation}
{\cal D}_{(\mu )}=e_{(\mu )}^{\nu }\partial _{\nu }-\frac{1}{2}S^{\nu \rho
}G_{\nu \rho \mu }\quad ,\quad G_{\nu \rho \mu }=e_{(\nu );\kappa }^{\sigma
}e_{(\rho )\sigma }e_{(\mu )}^{\kappa },  \label{Dcal}
\end{equation}
where $e_{(\mu )}^{\nu }$ is an orthonormal vierbein and the indices in the
brackets belong to the local tangent plane. Let us choose the following
representation for the complex vierbein
\begin{equation}
e_{(i)}^{k}=\kappa _{{\bf x}}\delta _{i}^{k}-iR^{-1}\varepsilon
_{ikl}x^{l}\quad ,\quad e_{(i)}^{0}=e_{(0)}^{i}=0\quad ,\quad e_{(0)}^{0}=1.
\label{4. 24}
\end{equation}
Then, using the equality $V_{ik}J_{k}=0$, the covariant generalization of
Eqs.~(\ref{bargwig1a})-(\ref{bargwig2a}) takes the form
\begin{equation}
\left( \partial _{t}+s^{-1}J_{i}e_{(i)}^{k}\partial _{k}-\frac{s+1}{R}
\right) \psi =V_{+i}e_{(i)}^{k}\partial _{k}\psi =0.  \label{4. 35}
\end{equation}
It can be shown~\cite{III} that at $\psi \sim e^{i\omega t}$ and $s=1/2,\ s=1$
the above equations correspond to the irreducible representations of the
Lorentz group with the weights~(\ref{weights}).

In the Minkowski space the solutions of Eqs.~(\ref{bargwig5}) are the
products of the plane waves $e^{-ip\cdot x}$ and  so-called massless
spinors for the finite-dimensional representations of the Lorentz group
constructed in~\cite{48}. Let us denote the stability subgroup
 of the Minkowski light-like vector $(1,0,0,1)$ as ${\cal H}$; this group is
called the little Lorentz group of this vector.  Let us consider the vector
in the  ${(j_{+},j_{-})}$ representation invariant under ${\cal H}$ with
respect to phase transformations. Let us act on this vector by the
transformation which turns the vector $(1,0,0,1)$ into the light-like vector
$n^{\mu }=(\omega ,\omega {\bf q})\ ,\ {\bf q}^{2}=1$.  The vector from  the
 space of the ${(j_{+},j_{-})}$ representation  corresponding to $n^{\mu }$
is just the spinor connected with the vector $n^{\mu }$. The above scheme
realizes the construction of CS for the  $SO(3,1)/{\cal H}$ space in
accordance with the general method proposed for an arbitrary Lie group in
\cite{coher1}. One can obtain the polynomial form of the above mentioned
spinors
\begin{equation}
|{\bf q}\omega ;j_{+}j_{-}\rangle =\left( \omega
\frac{1+q^{3}}{2}\right) ^{j_{+}+j_{-}}(1-z\rho _{{\bf
q}})^{2j_{+}}(1-\bar{z}\bar{\rho}_{{\bf q} })^{2j_{-}},  \label{lor6}
\end{equation}
where $\rho _{{\bf q}}=\frac{q^{1}+iq^{2}}{1+q^{3}}$. The spinor constructed
in~\cite{40} coincides with the above expression up to the factor $\omega
^{j_{+}+j_{-}}$.

It is easy to show that the equalities
\begin{eqnarray}
(s^{-1} q^i J_i - 1) |{\bf q}\omega;s0\rangle = V_{+i}q_i |{\bf q}
\omega;s0\rangle =0,  \label{4. 36} \\
(V_{+i}+iV_{+m}\varepsilon_{imn}q^n )|{\bf q}\omega;s0\rangle =0,
\label{4. 37} \\
(i\varepsilon_{imn}J_m q^n -sq^i + J^i ) |{\bf q}\omega;s0\rangle =0
\label{4. 38}
\end{eqnarray}
hold for any spin. Note that Eq.~(\ref{4. 38}) follows from the 
equation~(\ref{4.  36}) and decomposition~(\ref{bargwig5a}).

To generalize the solutions~(\ref{spin0}) for   arbitrary spin particles,
let us construct  the functions
\[
f_{{\bf q\omega }}^{(s)}({\bf x})=\phi _{{\bf q}}({\bf x};i\omega R-s-1)|
{\bf q}\omega ;s0\rangle .
\]
Using~(\ref{4. 36})-(\ref{4. 38}) we find that the functions
\begin{equation}
\psi ({\bf x},t)=e^{i\omega t}f_{{\bf q\omega }}^{(s)}({\bf x})
\label{4. 21}
\end{equation}
obey Eqs.~(\ref{4. 35}) for an arbitrary spin. Under $R\rightarrow \infty$
the functions~(\ref{4. 21}) pass into the usual plane waves in the Minkowski
space considered in \cite{40} up to the multiplier $\omega ^{s}$. The
solutions~(\ref{4. 21}) are much simpler than the ones obtained for particles 
of spin~0, 1/2 and~1 over ${\bf R}^{1}\times S^{3}$ and $S^{1}\times S^{3}$ 
spaces by other methods~\cite{15/42/72}.

Thus, like the case of de Sitter space~\cite{dS-PLB}, solutions of the
spinor wave equations over $H^3$ space are the product of  two CS systems
with different stability subgroups. Then we can write down
transformation properties of the  $f_{{\bf q\omega }}^{(s)}({\bf x}) functions
$ under the action of the Lorentz group and then construct the invariant
two-point functions over the $H^{3}$ space. Also we can construct quantized
massless fields over the ${\bf R}^{1}\times H^{3}$ space using the
functions~(\ref{4. 21}) and the creation-annihilation operators with the
canonical (anti)commutation relations. The propagators of these fields can
be found using the above mentioned two-point functions over the $H^{3}$
space. In the spin zero case the Lorentz-invariant scalar product of  two
CS is the two-point function~\cite{coher1}
\[
{\cal W}^{(0)}({\bf x},{\bf y};\omega )=\int_{S^{2}}d^{2}q\,f_{{\bf q\omega }
}^{(0)}({\bf x})f_{{\bf q,-\omega }}^{(0)}({\bf y}),
\]
where ${\bf x}$ and ${\bf y}$  arbitrary points of the $H^{3}$ space and
$d^{2}q=dq^{1}dq^{2}/q^{3}$ is the rotational-invariant measure over $S^{2}$.
Using the integral representations of the hypergeometric function~\cite{59}
we get
\begin{equation}
{\cal W}^{(0)}({\bf x},{\bf y};\omega )=4\pi \mathop{_2 F_1}\left( \frac{
-i\omega R+1}{2},\frac{i\omega R+1}{2},\frac{3}{2};-\sinh ^{2}\alpha \right)
,  \label{4. 32a}
\end{equation}
where $\alpha $ is the geodesic distance between the ${\bf x},{\bf y}$
points defined by the relation $\cosh \alpha =\kappa _{{\bf x}}\kappa _{{\bf
y}}-R^{-2}{\bf xy}.$

Now let us construct the quantized massless field
\[
\Psi (x)=\int_{0}^{\infty }\omega d\omega \int_{S^{2}}d^{2}q\,\left(
e^{i\omega x^{0}}f_{{\bf q\omega }}^{(0)}({\bf x})a({\bf n};s)+e^{-i\omega
x^{0}}f_{{\bf q,-\omega }}^{(0)}({\bf x})a^{\dagger }({\bf n};s)\right) ,
\]
where the bosonic creation-annihilation operators $a({\bf n})$ and
$a^{\dagger }({\bf n})$ obeying
\[
\lbrack a({\bf n}),a^{\dagger }({\bf n}^{\prime })]=n^{0}\delta ^{3}({\bf n}-
{\bf n}^{\prime })
\]
are introduced and $n^{\mu }=(\omega ,{\bf q}\omega ).$ Then the
corresponding propagator
\[
D(x,y)\equiv [\Psi (x),\Psi ^{\dagger }(y)]=\int_{-\infty }^{\infty }\omega
d\omega e^{i\omega (x^{0}-y^{0})}{\cal W}({\bf x},{\bf y};\omega )
\]
is invariant under the time translations and  spatial $SO(3,1)$
transformations. Using formulas~\cite{59} for the
hypergeometric function at the special values of parameters the expression~(
\ref{4. 32a}) can be represented as
\begin{equation}
{\cal W}({\bf x},{\bf y};\omega )=\frac{4\pi \sin \omega R\alpha }{\omega
R\sinh \alpha }.  \label{twop-h3}
\end{equation}
The representation~(\ref{twop-h3}) coincides with the spherical function for
the spin-zero infinite-dimensional representations of the Lorentz group~\cite
{49}. After simple manipulations with the $\delta $-function we obtain
\[
D({\bf x},t_{1};{\bf y},t_{1}+t)=-\frac{4\pi i\alpha }{\sinh \alpha }\delta
(t^{2}-R^{2}\alpha ^{2})\varepsilon (t).
\]
The above expression coincides with the difference of the positive- and
negative-frequency Wightmann functions, obtained previously by the standard
methods~\cite{33}, within the multiple $(4\pi )^{2}$.

I am grateful to Yu.P.Stepanovsky for helpful discussions and to
O.B.Zaslavskii for useful notes.

\end{document}